\documentclass[a4paper, 11pt]{amsart}

\usepackage{amsmath, amssymb, amsthm}
\usepackage[english]{babel}
\usepackage[T1]{fontenc}
\usepackage{enumerate}
\newcommand{\ee}{\mathbf{e}}

\def\N {\mathbb{N}}

\def \be {  \varpi}

\newcommand{\fer}[1]{(\ref{#1})}

\newcommand{\R}{\mathbb R}
\def\be#1\ee{\begin{equation}#1\end{equation}}

\numberwithin{equation}{section}

\newcommand{\bq}{\begin{equation}}
\newcommand{\eq}{\end{equation}}

\newtheorem{thm}{Theorem}

\theoremstyle{rem}
\newtheorem{rem}[thm]{Remark}
\theoremstyle{definition}

\newenvironment{equations}{\equation\aligned}{\endaligned\endequation}

\begin{document}

\title[Entropy-type inequalities for generalized Gamma densities]{ENTROPY-TYPE INEQUALITIES FOR  \\ GENERALIZED GAMMA DENSITIES}

\author{GIUSEPPE TOSCANI}
\address{Department of Mathematics, University of Pavia, and IMATI of the National Council for Research;
via Ferrata 1,
Pavia, 27100 Italy}
\email{giuseppe.toscani@unipv.it}

\maketitle

\begin{abstract}
We investigate the relaxation to equilibrium of the solution of a class of one-dimensional linear Fokker--Planck type equations that have been recently considered in connection with the study of addiction phenomena in a system of individuals. The steady states of these equations belong to the class of generalized Gamma densities. As a by-product of the relaxation analysis, we prove new  weighted Poincar\'e and logarithmic Sobolev type inequalities for this class of densities.
\end{abstract}


\section{Introduction}\label{intro}

Let $X$ be a random variable whose density function is
 \be\label{equili}
f_\infty(x;\theta, \kappa,\delta) = \frac \delta{\theta^\kappa} \frac 1{\Gamma\left(\kappa/\delta \right)} x^{\kappa -1}
\exp\left\{ - \left( x/\theta\right)^\delta\right\}.
 \ee 
for non-negative values of $x$ and positive values of the parameters $\kappa, \theta$ and $\delta$, denoting respectively the shape, the scale and the exponent. The function \fer{equili} has been considered as a generalization of the Gamma distribution by Stacy \cite{Sta}, and includes the familiar Gamma, Chi, Chi-squared, exponential and Weibull densities  as special cases. Generalized Gamma distributions, also known as Amoroso and Stacy-Mihram distributions \cite{Amo,JKB}, are widespread in physical and biological sciences, as well as in the field of social sciences. Among others, they  describe rainfall run-off from a watershed \cite{Li}, wind speeds distribution \cite{JHMG}, the occurrence of the failure of a component or system \cite{Lie}, or events history and survival analysis \cite{box}. 

In the field of social phenomena, generalized Gamma densities have been shown to characterize equilibrium densities in a kinetic model aiming to describe alcohol consumption in a multi-agent society \cite{DT}. The kinetic description in \cite{DT}, inspired by the fitting analysis of alcohol consumption in \cite{Keh, Reh}, has been subsequently generalized to other social phenomena of addiction, including gambling activity and the abuse of the insights of social networking sites \cite{To3}. 

In \cite{To3} the  kinetic modeling of the addiction phenomena  led to a Fokker--Planck type equation describing relaxation of the statistics of the underlying addiction towards a steady state. Let $f = f(x,t)$  denote the probability density of individuals which are characterized by  an addiction value  equal to $x\in \R_+$ at time $t\ge 0$.  The time-evolution of the density $f(x,t)$ was shown in \cite{To3} to obey to a linear Fokker--Planck type equation with variable coefficients of diffusion and drift, given by 
\begin{equation}\label{FP}
 \frac{\partial f(x,t)}{\partial t} =  \frac{\partial^2 }{\partial x^2}
 \left(x^{2-\delta} f(x,t)\right )+ 
 \frac{\partial}{\partial x}\left[ \left(  \frac\delta{\theta^\delta} x\, -(\kappa +1-\delta)x^{1-\delta}\right) f(x,t)\right].
 \end{equation}
In \fer{FP} $\theta$, $\kappa$ and $\delta$ are positive constants  related to the relevant features of the addiction phenomenon under study. The modeling assumptions in \cite{DT,To3} moreover imply the bound $\delta \le 1$. As shown in \cite{To3}, the Fokker--Planck type equation \fer{FP} has a unique equilibrium density, given by the generalized Gamma $f_\infty(x;\theta, \kappa,\delta)$ defined in \fer{equili}. This fact establishes a strict link between addiction phenomena in a multi-agent society and generalized Gamma densities.

One of the physically relevant problems related to the Fokker--Planck equation \fer{FP} is the study of the exact rate of relaxation to equilibrium of its solution. Indeed, as it happens for the classical Fokker--Planck equation \cite{To99}, an exponential in time rate of relaxation justifies the fact that in addiction phenomena we are essentially observing a generalized Gamma equilibrium density \cite{Keh,Reh}. This allows to consider equation \fer{FP} as a reasonable model for the statistical description of the phenomenon under study \cite{FPTT}.

The strategy used for the classical Fokker--Planck equation, corresponding to constant coefficient of diffusion and linear drift, suggests to study the relaxation of the solution of \fer{FP} towards equilibrium by looking at the time evolution of its Shannon entropy relative to the equilibrium density. We recall that, given two probability densities $f(x)$ and $g(x)$, with $x \in \R_+$, the Shannon entropy of $f$ relative to  $g$ is defined by
 \be\label{rel-H}
H(f|g) = \int_{\R_+} f(x) \log \frac{f(x)}{g(x)}\, dx.
 \ee
In this note, we will prove that the rate of relaxation in relative Shannon entropy is exponential in time, in reason of a new weighted logarithmic Sobolev inequality satisfied by the generalized Gamma density. 

The deep link between differential inequalities of Sobolev type and Fokker--Planck  equations has been first remarked for the classical Fokker--Planck equation. While the standard logarithmic Sobolev inequality allows to prove exponential convergence of the solution towards the Maxwellian equilibrium density in relative entropy, at the same time the evolution of the relative entropy of the solution density of the Fokker--Planck equation can be used to obtain a dynamical proof of the logarithmic Sobolev inequality  \cite{To97,To99}. This idea has been subsequently extended, to obtain sharp differential inequalities,  to  Fokker--Planck type equations with constant diffusion term and general drift by Otto and Villani \cite{OV}. Then, a variant of the method has been recently applied to Fokker--Planck type equations with variable coefficient of diffusion arising in social and economical applications \cite{FPTT19, FPTT20}. Also, Fokker--Planck type equations with variable coefficient of diffusion and linear drift have been shown to be a useful instrument to prove weighted Poincar\'e type inequalities \cite{FPTT}. In the present case, one has to remark that the underlying Fokker-Planck equation \fer{FP} does not have a linear drift (except when $\delta =1$), and the result of \cite{FPTT} can not be applied directly.

The forthcoming analysis will take advantage of a special property of the Fokker--Planck type equations \fer{FP}, which are shown to be linked each other for different values of the parameter $\delta$. By means of this property, we can directly prove existence and uniqueness of the solution for the whole set of parameters, by resorting to the results of a pioneering paper by Feller \cite{Fel1}, which refers to the qualitative analysis of the case $\delta =1$. This will be done in the next Section \ref{analisi}. Then, the main weighted differential inequalities for the generalized Gamma densities will be obtained in Section \ref{dis}. These results allow to compute in Section \ref{rate}, with an explicit rate, the convergence towards equilibrium of the relative entropy of solution to the Fokker--Planck equation \fer{FP}.

\section{A qualitative study of the Fokker--Planck equation}\label{analisi}

Before studying the rate of relaxation to equilibrium of the solution to the Fokker--Planck type equation \fer{FP}, we will give results on existence and uniqueness of solutions to its initial-boundary value problem.  As extensively discussed in \cite{FPTT,PT13}, the natural boundary conditions associated to equation \fer{FP} are the so-called \emph{no--flux} expressed by 
\be\label{bc1}
\left. \frac{\partial }{\partial x}\left( x^{2-\delta} f(x,t)\right) + \left(  \frac\delta{\theta^\delta} x\, -(\kappa +1-\delta)x^{1-\delta}\right) \,f(x,t) \right|_{x=0,+\infty} = 0, \quad t>0.
\ee
Conditions \fer{bc1} guarantee, at least formally,   the correct evolution of the main macroscopic quantities of the solution to  equation \fer{FP}, and among them the mass conservation. 

To be precise, let $\mathcal{M}_0$ define the space of all probability measures in $\R_+$, and for $\alpha\ge 0$, let  
\be\label{ma}
\mathcal{M}_\alpha= \left\{ \mu \in \mathcal{M}_0: \int_{\R_+} x^\alpha \mu(dx) <+\infty\right\}
\ee
denote the space of all Borel probability measures of finite moment of order $\alpha$, equipped with the topology of the weak convergence of measures. 
In presence of the no-flux boundary conditions \fer{bc1} one can study, without loss of generality,  the initial-boundary value problem for equation \fer{FP} with an initial datum $f_0(x) \in \mathcal{M}_\alpha$, for some $\alpha \ge 0$. Then, since \fer{bc1} imply mass conservation, it follows that the solution $f(x,t) \in \mathcal{M}_0$  for all subsequent times $t >0$.

\begin{rem}
In consequence of the modeling assumptions (cf. the discussion in \cite{DT,To3}) the values of the constant $\delta$ in the addiction phenomena are restricted to the interval $0 < \delta \le 1$. However, as far as the Fokker--Planck type equation \fer{FP} is concerned, we will consider in the following values of $\delta$ in the interval $0< \delta \le 2$, thus covering generalized Gamma densities that range from the Log-normal density, corresponding to $\delta \to 0$, to Chi-densities, obtained for $\delta =2$.
\end{rem} 

 Fokker--Planck type equations of type \fer{FP} can be rewritten in different equivalent forms, each one useful for various purposes. The most used alternative form (its adjoint) \cite{Fel2,FPTT,OV}, is obtained by looking at the evolution of the function 
\be\label{ad-v}
h(x,t) = \frac{f(x,t)}{f_\infty(x;\theta,\kappa,\delta)}.
\ee
The new unknown function $h(x,t)$ satisfies the drift-diffusion 
equation
\begin{equation}\label{FPadj}
 \frac{\partial h(x,t)}{\partial t} =  x^{2-\delta}\, \frac{\partial^2 }{\partial x^2}
  h(x,t)-  \left(  \frac\delta{\theta^\delta} x\, -(\kappa +1-\delta)x^{1-\delta}\right) 
 \frac{\partial}{\partial x}h(x,t).
 \end{equation}
 
We will consider here a third version of the Fokker--Planck equations \fer{FP}, that highlights an interesting feature of their solutions. 
 For given $t >0$, let $X(t)$ denote the random process with probability density $f(x,t)$, solution of the Fokker--Planck equation \fer{FP}, and let
 \be\label{dist}
 F(x,t) = P(X(t) \le x)=  \int_0^x f(y,t)\, dy
 \ee
denote its probability distribution. 
Integrating both sides of equation \fer{FP} on the interval $(0,x)$, and applying condition \fer{bc1} at the boundary point $x=0$, it follows by simple computations that $F(x,t)$ satisfies the equation 
\begin{equation}\label{FP3}
 \frac{\partial F(x,t)}{\partial t} =  x^{2-\delta}\, \frac{\partial^2 }{\partial x^2}
  F(x,t) + \left(  \frac\delta{\theta^\delta} x\, -(\kappa -1)x^{1-\delta}\right) 
 \frac{\partial}{\partial x}F(x,t),
 \end{equation}
The no-flux boundary conditions \fer{bc1} then guarantee that, for any $t \ge 0$  
 \be\label{limo}
 F(0,t) = 0; \qquad \lim_{x \to +\infty} F(x,t) = 1,
 \ee
where the second condition in \fer{limo}  corresponds to mass conservation. 
 
Given a strictly positive constant $m$, let us consider the transformation 
\be\label{key}
 F(x,t) = G(y,\tau), \qquad y =y(x) = x^m, \quad \tau = \tau(t) = m^2 t.
\ee
Then it holds 
 \[
 \frac{\partial}{\partial t}F(x,t) = m^2 \,\frac{\partial}{\partial \tau}G(x,\tau),   
  \]
while
 \[
 \frac{\partial}{\partial x}F(x,t) =  m x^{m-1}\, \frac{\partial}{\partial y}G(y,\tau), 
 \]
and
\[
\frac{\partial^2}{\partial x^2}F(x,t) =  m^2x^{2m-2}\, \frac{\partial^2}{\partial y^2}G(y,\tau) + m(m-1) x^{m-2}\, \frac{\partial}{\partial y}G(y,\tau).
\]
Hence, substituting into \fer{FP3} the above identities and using the inverse relation $x = y^{1/m}$, one obtains that $G(y, \tau)$ satisfies the equation
\begin{equation}\label{FP4}
 \frac{\partial G(y,\tau)}{\partial \tau} =  y^{2-\delta/m}\, \frac{\partial^2 }{\partial y^2}
  G(y,\tau) + \left( \frac{\delta/m}{(\theta^m)^{\delta/m}} y\, -\left(\frac\kappa{m} -1\right)y ^{1-\delta/m}\right) 
 \frac{\partial}{\partial y} G(y,\tau).
 \end{equation}
 Moreover, if $F(x,t)$ satisfies conditions \fer{limo} for any $t \ge0$, $G(y,\tau)$ still   satisfies the same conditions  for any $\tau \ge 0$. Note that, like in the previous case, equation \fer{FP4} is obtained  from the Fokker--Planck equation
 \be\label{FP-m}
 \frac{\partial g(y,\tau)}{\partial \tau} =   \frac{\partial^2 }{\partial y^2}\left(y^{2-\delta/m}\,
  g(y,\tau)\right) + \frac{\partial}{\partial y} \left[  \left( \frac{\delta/m}{(\theta^m)^{\delta/m}} y\, -\left(\frac\kappa{m} +1 -\frac\delta{m}\right)y ^{1-\delta/m}\right) g(y,\tau)\right],
 \ee
 integrating both sides of equation \fer{FP-m} on the interval $(0,y)$, and applying condition \fer{bc1} at the boundary point $y=0$.
Note moreover that equation \fer{FP-m} has the same structure of equation \fer{FP}, with the constants $\kappa$,  $\theta $, and  $\delta $ substituted by $\theta^m$, $\kappa/m$  and  $\delta/m$ . Consequently, its equilibrium distribution is given by the generalized Gamma density 
 \be\label{equi2}
 f_\infty\left(y;\theta^m, \frac\kappa{m},\frac\delta{m}\right).
 \ee
It is interesting to remark that, if $F(x,t)$ is the  probability distribution of the random process $X(t)$, the process $Y(\tau)$ with probability distribution $G(y,\tau)$ has the same law of the process $X^m(\tau/m^2)$. Indeed, thanks to transformation \fer{key}
 \begin{equations}\label{scala}
P(Y(\tau) \le y)& = G(y,\tau) =  F\left(y^{1/m},\frac\tau{m^2}\right) = P\left( X\left(\frac\tau{m^2}\right) \le y^{1/m}\right) =\\
 & P\left( X^m \left(\frac\tau{m^2}\right) \le y\right).
 \end{equations}
In  \cite{Sta}, Stacy noticed that the generalized Gamma densities satisfy a similar property. Given a constant $m >0$, when a random variable is distributed according to \fer{equili},  $X^m$ is distributed according to \fer{equi2}. 

The previous result can be rephrased in the following way.  Given a random variable $X_0$ with probability density $f_0(x)$,  and denoting by $X(t)$ the process with probability density given by the  solution $f(x,t)$ to the initial-boundary value problem \fer{FP}-\fer{bc1}, for any choice of the constant $m>0$  the solution to the new initial-boundary value problem for the Fokker--Planck equation \fer{FP} with parameters $\theta^m$, $\kappa/m$ and $\delta/m$ departing from the density of $X_0^m$ is given by the density of the process $X^m(t/m^2)$. 

Thus, with respect to the parameter $m >0$, the Fokker--Planck type equations \fer{FP}
are connected each other through the scaling \fer{key}. Therefore, starting from equation \fer{FP}, by choosing $m$ in the interval $[\delta/2, +\infty)$, one obtains as equilibrium densities all generalized Gamma densities with exponents in the interval $(0,2]$, being the exponent $2$ (the Chi-density) realized for $m = \delta/2$, and the exponent $0$ (the Log-normal density \cite{Aic}) achieved in the limit $m \to +\infty$ when, for any given $\delta >0$, $\theta,\kappa$ are given by \cite{DT}
 \be\label{para}
\theta =  \left(\frac\delta{m}\right)^{2/\delta}, \quad  \kappa =  m\left( \frac{m}\delta + \frac \delta{m} -1\right).
 \ee
The aforementioned discussion allows to conclude that, once a result for the initial-boundary value problem for the Fokker--Planck type equation \fer{FP} has been achieved  for a certain value $\delta = \delta_0$ of the exponent, the same result holds for all the other values $\delta\not=\delta_0$ of the interval $0<\delta \le 2$. 

As we shall see, to easily obtain the desired results, the best choice is to restrict the study to two special cases, corresponding to the values $m =\delta$ and $m =\delta/2$. Inserting these values of $m$ into the Fokker--Planck equation \fer{FP-m} leads to simplify the drift term and, respectively,  the diffusion coefficient. The choice $m =\delta$  leads to a Fokker--Planck equation with linear drift
\begin{equation}\label{FP-1}
 \frac{\partial f(x,t)}{\partial t} =  \frac{\partial^2 }{\partial x^2}
 \left(x f(x,t)\right )+ 
 \frac{\partial}{\partial x}\left[ \left(  \frac x{\theta^\delta} \, -\frac\kappa\delta \right) f(x,t)\right].
 \end{equation}
Likewise, setting $m=\delta/2$ in \fer{FP-m} leads to a Fokker--Planck equation with constant coefficient of diffusion given by
\begin{equation}\label{FP-2}
 \frac{\partial f(x,t)}{\partial t} =  \frac{\partial^2 }{\partial x^2} f(x,t)+ 
 \frac{\partial}{\partial x}\left[ \left(  \frac2{\theta^\delta} x\, -\left(\frac {2\kappa}\delta -1\right)x^{-1}\right) f(x,t)\right].
 \end{equation}

Parabolic equations like \fer{FP-1} have been exhaustively studied in a pioneering paper by Feller  \cite{Fel1} (cf. also \cite{Fel2}). The class of parabolic equations studied in \cite{Fel1} is
\begin{equation}\label{Fel}
 \frac{\partial u(x,t)}{\partial t} =  \frac{\partial^2 }{\partial x^2} (axu(x,t)) -
 \frac{\partial}{\partial x} \left[\left(bx+c\right)  u(x,t)\right],
 \end{equation}
where $x \in \R_+$, $a, b$ and $c$ are constants, and $a > 0$.  Equation \fer{Fel} in \cite{Fel1} has been coupled with a boundary condition at $x=0$. In the notation of \cite{Fel1} the quantity
 \be\label{flux}
 \mathcal{F}(t) = \lim_{x \to 0}  \left\{ -\frac{\partial }{\partial x} (axu(x,t)) +
 \left(bx+c\right)  u(x,t)\right\}
 \ee 
is called the flux of $u$ at the origin. Then, condition \fer{bc1} corresponds to assume $\mathcal F (t) = 0$ identically in time.
In \cite{Fel1}, it was proven that in most cases the solutions corresponding to the no-flux boundary conditions \fer{bc1}  would preserve positivity and norm, so that
\[ 
\int_{\R_+} u(x,t) \, dx = \int_{\R_+} u(x,t=0) \, dx.
\]
In particular, if $0 < c < a$ there exists  a unique solution of the initial-boundary value problem (that is a positivity and norm preserving solution) defined by the condition that $u_x(x,t)$ at $x = 0$ vanishes, while
if $c > a$ there exists a unique positivity and norm preserving solution of the initial-boundary value problem such that both it and $u_x(\cdot,t)$ vanish at $x = 0$. This means that when $c>a$ the boundary $x = 0$ acts both as absorbing  and reflecting barrier and that no homogeneous boundary conditions can be imposed. The solutions in \cite{Fel1} are functions $u(x,t)\in L_1(\R_+)$  such that, for $x>0$ have continuous partial derivatives satisfying \fer{Fel}, and, for every fixed $s>0$ and $t>0$ the functions $e^{-sx}u(x,t)$ and $e^{-sx}u_t(x,t)$ are integrable over $0<x<+\infty$, and this uniformly in every interval $0<t_0 \le t \le t_1 < +\infty$.

In case of $L_1$-functions, uniqueness can be easily proven also by resorting  to a classical argument \cite{Zua}, which implies that the $L_1$-norm of the difference of two solutions is not increasing in time. 

Another important property of the solutions to equation \fer{Fel}, not investigated by Feller, concerns the evolution in time of the principal moments
 \[
 M_n(t) = \int_{\R_+} x^n\, u(x,t) \, dx, \qquad n \in \N_+.
 \]
Resorting to a recursive argument, it is a simple exercise to verify that, provided $b <0$, the boundedness of the $n$-th moment $M_n(t=0)$ implies the uniform boundedness of the $n$-th moment at any subsequent time $t>0$.  

Since the Fokker--Planck equation \fer{FP-1} is of type \fer{Fel}, the results of \cite{Fel1} guarantee that, for any given initial data $f_0(x)\in \mathcal{M}_\alpha$, $\alpha \ge 0$ we have two different existence and uniqueness results. When $0 <\kappa<\delta$ and in presence of the no-flux boundary conditions \fer{bc1}, there exists a unique positive and norm preserving solution of the initial-boundary value problem. If  $\kappa >\delta$,  there exists a unique positive and norm preserving solution of the initial-boundary value problem such that both it and its flux vanish at $x = 0$. This means that  when  $\kappa >\delta$ the boundary $x = 0$ acts both as absorbing and reflecting barrier and that no homogeneous boundary conditions need to be imposed. Mass conservation holds even without no flux boundary conditions. 
In both cases, if the initial value $f_0 \in \mathcal{M}_\alpha$ for some $\alpha >0$, the solution $f(\cdot, t) \in \mathcal{M}_\alpha$  uniformly  in time. These solutions satisfy the regularity property quoted above for the solutions to the parabolic equation \fer{Fel}.

Taking into account the connection among the Fokker--Planck equations \fer{FP} and \fer{FP-m}, the existence and uniqueness results relative to the case $\delta =1$ still hold for the initial-boundary value problem for equation \fer{FP} characterized by a parameter $\delta \not=1$. For a given initial probability density $f_0\in \mathcal{M}_\alpha$, $\alpha \ge 0$, and in presence of boundary conditions like the ones given in \fer{bc1}, there exists a unique positive and mass preserving solution $f(\cdot,t) \in \mathcal{M}_\alpha$. Moreover,  if $\kappa > \delta$,  there exists a unique positive and norm preserving solution of the initial value problem such that both it and its flux vanish at $x = 0$.  Mass conservation holds in this range of the parameters even without no flux boundary conditions.

\begin{rem} The limit case $\delta \to 0$, with $\theta,\kappa$ given as in \fer{para} with $m=1$, leads to the Log-normal equilibrium density
 \be\label{LN}
 f_{LN}(x) = \frac 1{\sqrt{2\pi}\,x}\exp\left\{-\frac 12\left( \log x +1\right)^2 \right\}, 
 \ee
 steady state solution of the Fokker-Planck equation derived and studied in \cite{GT18, GT19}
\begin{equation}\label{FPLN}
 \frac{\partial f(x,t)}{\partial t} =  \frac{\partial^2 }{\partial x^2}
 \left(x^2 f(x,t)\right )+ 
 \frac{\partial}{\partial x}\left[\left( x\,log x\right) f(x,t)\right].
 \end{equation}
Hence it follows that in this case no homogeneous boundary conditions need to be imposed on equation \fer{FPLN} to have a unique positive and mass preserving solution. 
 \end{rem}

\section{Differential inequalities for generalized Gamma densities}\label{dis}

 The aim of this Section is to prove that the class of densities \fer{equili}, steady state solutions of the Fokker--Planck type equations \fer{FP}, satisfy some weighted inequalities of  Poincar\'e and logarithmic Sobolev type.

Let $X$ be a random variable distributed with probability density $f(x)$, where $x\in \R_+$. The random variable $X$ is said to satisfy a weighted Poincar\'e-type inequality with weight function $\lambda(x)$ (where $\lambda$ is a fixed nonnegative, Borel measurable function), if for any bounded smooth function $\phi$ on $\R_+$ 
 \be\label{Poi}
 Var\left[\phi(X)\right] \le E\left\{\lambda(X)[\phi'(X)]^2\right\}
 \ee
 As usual
 \[
 Var\left[\phi(X)\right] = \int_{\R_+} \phi^2(x)\, f(x) \, dx - \left(\int_{\R_+} \phi(x)\, f(x) \, dx \right)^2
 \]
 stands for the variance of $\phi$ under $f$.  
 Likewise, $X$ is said to satisfy a weighted logarithmic Sobolev inequality with weight function $\lambda(x)\ge 0$ if, for any bounded smooth function $\phi$ on $\R_+$ 
 \be\label{PoiS}
 Ent\left[\phi^2(X)\right] \le E\left\{\lambda(X)[\phi'(X)]^2\right\}
 \ee
 Here
 \[
 Ent \left[\phi^2(X)\right] = \int_{\R_+} \phi^2(x)\log \phi^2(x)\, f(x) \, dx - \int_{\R_+} \phi^2(x)\, f(x) \, dx \, \log \int_{\R_+} \phi^2(x)\, f(x) \, dx 
 \]
 denotes the entropy of $\phi^2$ under $f$. 
 The inequalities are understood in the following sense: if the right-hand side is finite, then  the inequalities hold true.

Abstract weighted Poincar\'e and logarithmic Sobolev inequalities are connected with the problem of large deviations of Lipschitz functions and measure concentration. In reason of that, the question whether a probability measure satisfies such functional inequalities has attracted a lot of attention in recent years  \cite{BCG,BL,BJ,BJM1,BJM2,CGGR,Goz}. 

In the probabilistic literature, inequality \fer{Poi} is also known under the name of weighted Chernoff inequality, in reason of the analogous inequality with weight $\lambda(x) =1$ obtained by Chernoff \cite{Che} for the one-dimensional Gaussian density 
 \be\label{Max}
 g(x) = \frac 1{\sqrt{2\pi}} \exp\left\{-\frac{x^2}2\right\}, \qquad x \in \R.
 \ee
Chernoff-type inequalities with weight were proven, few years later Chernoff's result, by Klaassen \cite{Kla}, who listed a number of probability densities for which the weight $\lambda(x)$ was explicitly computable. Among others, Klaassen proved that a weighted Chernoff-type inequality holds for the Gamma density,  namely density \fer{equili} in which $\delta =1$ 
 \be\label{gam}
 f_\infty(x;\theta, \kappa,1) = \frac 1{\theta^\kappa} \frac 1{\Gamma\left(\kappa \right)} x^{\kappa -1}
\exp\left\{ - \left( x/\theta\right)\right\}.
 \ee
For the Gamma density \fer{gam}, Klaassen obtained the weight $\lambda(x) = \theta x$. Hence, if $X_{\theta,\kappa}$ is a random variable distributed with the Gamma density  \fer{gam}, for all smooth functions $\phi$ with finite variance it satisfies the following Chernoff inequality with weight
 \be\label{Poi-G}
 Var\left[\phi(X_{\theta,\kappa})\right] \le \theta E\left\{X_{\theta,\kappa}[\phi'(X_{\theta,\kappa})]^2\right\}.
 \ee
 Note that the weight does not depend on the value of the shape $\kappa$.
The same result was obtained in \cite{FPTT}, resorting to a simpler proof that is closely related to the Fokker--Planck description of the Gamma density. In \cite{FPTT} it was proven that, 
if $X$ is a random variable distributed with density $f_\infty(x)$, $x \in I \subseteq \R$, and $f_\infty$ satisfies the differential equality
\be\label{staz-2}
\frac{\partial }{\partial x}\left(\lambda(x)  f_\infty(x) \right) + (x -M)\,f_\infty(x) = 0, \quad x\in I, 
 \ee
where $\lambda(x) \ge 0$ on $I$, and the constant $M\ge 0$,  then for any smooth function $\phi$  defined  on $I$ such that $\phi(X)$ has finite variance  
\be\label{chernoff-gen}
 Var[\phi(X)] \le E\left\{\lambda(X)[\phi'(X)]^2\right\}
\ee
 with equality if and only if $\phi(X)$ is linear in $X$. The expression on the left-hand side of in \fer{staz-2} coincides with the flux of the Fokker--Planck equation
\begin{equation}\label{FPP}
 \frac{\partial f(x,t)}{\partial t} =  \frac{\partial^2 }{\partial x^2}
 \left(\lambda(x) f(x,t)\right )+ 
 \frac{\partial}{\partial x}\left[ \left( x-M \right) f(x,t)\right],
 \end{equation}
and, consequently, the differential equation \fer{staz-2}  identifies the steady states of \fer{FPP}. If we now consider the Fokker--Planck equation \fer{FP} in which $\delta =1$ (the Gamma case), the flux is given by
\be\label{staz-g}
\frac{\partial }{\partial x}\left( x\,f(x)\right) + \left(\frac x\theta -\kappa\right)\,f(x), \quad x\in \R_+,
\ee
and consequently, the Gamma density satisfies \fer{staz-2} with  $\lambda(x) = \theta x$ and $M=\kappa\theta$. This proves \fer{Poi-G} with the same weight obtained by Klaassen \cite{Kla}.  Inequality \fer{Poi-G} can be used to cover all the other values of $\delta \not= 1$, by resorting to the scaling property of the generalized Gamma densities, expressed by \fer{equi2}. If the random variable $Y$ is distributed with density \fer{equili}, $Y^\delta$ is Gamma distributed with shape $\kappa/\delta$ and scale $\theta^\delta$. Therefore $Y^\delta$ satisfies inequality \fer{Poi-G} with weight $\theta^\delta \,x$, so that, for any absolutely continuous function $\phi$ it holds
  \be\label{Poi-1}
 Var\left[\phi(Y^\delta)\right] \le \theta^\delta E\left\{Y^\delta[\phi'(Y^\delta)]^2\right\}.
 \ee
If we now set $\psi(x) = \phi(x^\delta)$, that implies
 \[
 \left[\phi'(x)\right]^2 = \frac 1{\delta^2 x^{2\delta -2}}\left[\psi'(x)\right]^2,
 \]
inequality \fer{Poi-1} leads to the following

\begin{thm}\label{cher}
Let  $Y= X_{\theta,\kappa,\delta}$ be a random variable of  density \fer{equili}, a generalized Gamma density of parameters $\kappa, \theta$ and $\delta$, with $0<\delta \le 2$. Then, for any bounded smooth function $\phi$, $X$ satisfies the weighted Chernoff (Poincar\'e) inequality
  \be\label{Poi-GG}
 Var\left[\psi(Y)\right] \le \frac{\theta^\delta}{\delta^2} E\left\{Y^{2-\delta}[\psi'(Y)]^2\right\}.
 \ee
 The weight function in \fer{Poi-GG} does nor depend on the shape $\kappa$.
 \end{thm}
 
\begin{rem}\label{rem-LN}The Log-normal equilibrium density \fer{LN} is obtained from the generalized Gamma density by setting $\theta$ as in \fer{para}, $m=1$ and letting $\delta \to 0$. In this case, the coefficient $\theta^\delta/{\delta^2} =1$.
 Consequently, if the random variable $X_{LN}$ is distributed according to \fer{LN}, it satisfies the Chernoff inequality with weight 
  \be\label{Poi-LN}
 Var\left[\psi(X_{LN})\right] \le  E\left\{X_{LN}^{2}[\psi'(X_{LN})]^2\right\}.
 \ee
Clearly, \fer{Poi-LN} can be directly obtained from the original Chernoff inequality \cite{Che}.

On the other side, if $\delta =2$, the weight function $\lambda(x)$ takes the constant value $\theta^2/4$. Hence, if the variable $X_C$ is Chi-distributed with scale coefficient $\theta$, for any value of the shape $\kappa$ it holds 
\be\label{Poi-chi}
 Var\left[\psi(X_C)\right] \le \frac{\theta^2}{4} E\left\{[\psi'(X_C)]^2\right\}.
 \ee
\end{rem}

Weighted logarithmic Sobolev inequalities for generalized Gamma densities can be similarly derived by starting from the Fokker--Planck equation \fer{FP-2}, which is characterized by a constant diffusion coefficient.
Fokker--Planck equations  with constant diffusion and general drift have been considered by Otto and Villani \cite{OV} as a key instrument to obtain various differential inequalities. Their results can be directly applied to our case, to show that, provided $\kappa \ge \delta/2$, the generalized Gamma densities \fer{equili} satisfy the weighted logarithmic Sobolev inequality
 \be\label{LS-G}
  H(f|f_\infty(\theta, \kappa,\delta)) \le \frac{\theta^\delta}{\delta^2} I_{2-\delta}(f|f_\infty(\theta, \kappa,\delta)),
 \ee
where, given two probability densities $f(x)$ and $g(x)$, with $x \in \R_+$, $H(f,g)$ denotes  the Shannon entropy of $f$ relative to  $g$ defined in \fer{rel-H},
while, for a given constant $\beta \ge 0$,  $I_\beta(f,g)$ denotes the weighted Fisher information of $f$ relative to $g$
 \be\label{fis}
 I_\beta(f|g) = \int_{\R_+} x^\beta f(x) \left(\frac d{dx}\log \frac{f(x)}{g(x)}\right)^2\, dx.
 \ee
 In the notations of \cite{OV}, the Fokker--Planck equation \fer{FP-2} can be rewritten as
 \begin{equation}\label{FP-OV}
 \frac{\partial h(x,t)}{\partial t} =  \frac{\partial^2 }{\partial x^2} h(x,t)+ 
 \frac{\partial}{\partial x}\left(w'(x) h(x,t)\right), \quad x \in  I \subseteq \R.
 \end{equation}
 The potential $w(x)$ is such that the equilibrium solution of the Fokker--Planck equation \fer{FP-OV}, given by
 \be\label{eq}
 h_\infty(x) = C \, e^{-w(x)}. 
 \ee
can be made a probability density for a suitably chosen constant $C>0$. Then, if the potential $w(x)$ is uniformly convex on $I$, and
 \be\label{uni}
 w''(x) \ge \rho,
 \ee
it is proven in \cite{OV} that $h_\infty$ satisfies the logarithmic Sobolev inequality 
 \be\label{LS-L}
  H(h|h_\infty) \le \frac{1}{2\rho} I_0(h|h_\infty).
 \ee
For the Fokker--Planck equation \fer{FP-2} we have the identity
 \[
 w'(x) =  \frac2{\theta^\delta} x\, -\left(\frac {2\kappa}\delta -1\right)x^{-1}, \qquad x \in \R_+,
 \]  
and the steady state is the Chi-distribution $h_\infty = f_\infty(\cdot;\theta, \kappa,2)$.  Provided $\kappa \ge \delta/2$, $w(x)$ is uniformly convex on $\R_+$ and
  \[
  w''(x) \ge \frac2{\theta^\delta}.
    \]
 Consequently, the Chi-distribution $h_\infty$ satisfies the logarithmic Sobolev inequality 
 \be\label{LS-LN}
  H(h|h_\infty) \le \frac{\theta^\delta}{4} I_0(h|h_\infty).
 \ee
Inequality \fer{LS-LN} allows to show that  the Chi-distribution $h_\infty$ satisfies \fer{PoiS} (with constant weight) simply recalling that the relative Fisher information $I_0$ on the right-hand side of \fer{LS-LN} can be rewritten in a different way \cite{FPTT,JB}. It holds
  \be\label{diff}
 I_0(h|h_\infty) = 4 \int_{\R_+} h_\infty(x) \left( \frac d{dx}\sqrt{ \frac{h(x)}{h_\infty(x)}}\right)^2 \, dx
  \ee
Let $\phi(x)$  denote the function
 \[
 \phi(x) = \sqrt{ \frac{h(x)}{h_\infty(x)}},
  \]
 which is such that
  \be\label{norm}
\int_{\R_+} \phi(x)^2 h_\infty(x) \, dx = 1.
  \ee
Then, if $X$ is distributed according to  $h_\infty$, and $\kappa \ge \delta/2$, inequality \fer{LS-LN} can be rewritten as the logarithmic Sobolev inequality \fer{PoiS}, where $\lambda (x) = \theta^\delta$
\be\label{P-chi}
 Ent\left[\phi^2(X)\right] \le \theta^\delta E\left\{(\phi'(X))^2\right\},
 \ee
 since, in view of condition \fer{norm}, it holds the identity
 \[
 H(h|h_\infty) =  Ent\left[\phi^2(X)\right].
 \]
Inequality \fer{P-chi} remains clearly valid for any other function $\phi(x)$ that does not satisfy condition \fer{norm}. 

As for the case of the Chernoff inequality with weight, inequality \fer{Poi-chi} can be used to obtain analogous inequality for all the other values of $\delta <2$, by resorting to the scaling property of the generalized Gamma densities, expressed by \fer{equi2}. If the random variable $Y$ is distributed with density \fer{equili}, $Y^{\delta/2}$ is Chi-distributed with shape $2\kappa/\delta$ and scale $\theta^\delta$. Therefore $Y^{\delta/2}$ satisfies inequality \fer{Poi-G} with weight $\theta^\delta$. Hence, for any absolutely continuous function $\phi$ it holds
  \be\label{Poi-11}
 Ent\left[\phi(Y^{\delta/2})\right] \le \theta^\delta E\left\{[\phi'(Y^{\delta/2})]^2\right\}.
 \ee
If we now set $\psi(x) = \phi(x^{\delta/2})$, that implies
 \[
 \left[\phi'(x)\right]^2 = \frac 4{\delta^2 x^{\delta -2}}\left[\psi'(x)\right]^2,
 \]
inequality \fer{Poi-chi} leads to the the following

\begin{thm}\label{logS}
Let  $Y= X_{\theta,\kappa,\delta}$ be a random variable of  density \fer{equili}, a generalized Gamma density of parameters $\kappa, \theta$ and $\delta$, with $0<\delta \le 2$. Then, provided $\kappa \ge \delta/2$, for any bounded smooth function $\phi$ the random variable $Y$ satisfies the weighted logarithmic Sobolev inequality
 \be\label{Poi-G2}
 Ent\left[\psi(Y)\right] \le \frac{4\, \theta^\delta}{\delta^2} E\left\{Y^{2-\delta}[\psi'(Y)]^2\right\}.
 \ee
 The weight function in \fer{Poi-G2} does not depend on the shape $\kappa$.
 \end{thm}

\begin{rem} Resorting to the arguments of Remark \ref{rem-LN}, we obtain that, if the random variable $X_{LN}$ is distributed according to the Log-normal density \fer{LN}, it satisfies the weighted logarithmic Sobolev inequality
  \be\label{Poi-LN2}
 Ent\left[\psi(X_{LN})\right] \le  4\, E\left\{X_{LN}^{2}[\psi'(X_{LN})]^2\right\}.
 \ee
\end{rem}

\section{Exponential in time convergence to equilibrium}\label{rate}

Let us suppose that $H(f(t=0)|f_\infty(\theta, \kappa,\delta))$, namely the entropy of the initial value of the Fokker--Planck equation \fer{FP} relative to its equilibrium density \fer{equili} is bounded, and let us study the evolution in time of  $H(f(t)|f_\infty(\cdot;\theta, \kappa,\delta))$, where $f(t)$ is the solution to \fer{FP}. Repeating step-by step the computations of Section 3 in \cite{FPTT} one shows that
 \be\label{decc}
 \frac d{dt} H(f(t)|f_\infty(\theta, \kappa,\delta)) = - I_{2-\delta}(f(t)|f_\infty(\theta, \kappa,\delta)).
 \ee
where $I_{2-\delta}(f|g)$ is the relative Fisher information defined in \fer{fis}. We remark that the computations leading to \fer{decc} are rigorously justified by the regularity properties of the solution to \fer{FP}, and by its behavior on the boundaries (cf. the results in \cite{Fel1}). Inequality \fer{Poi-G2} of Theorem \ref{logS}  then implies the bound
 \be\label{bound}
 I_{2-\delta}(f(t)|f_\infty(\theta, \kappa,\delta)) \ge \frac{\delta^2}{\theta^\delta} H(f(t)|f_\infty(\theta, \kappa,\delta)).
 \ee
Making use of inequality \fer{bound} into \fer{decc} implies the exponential in time convergence of the relative entropy of the solution to the Fokker--Planck equation \fer{FP}  with an explicit rate. It holds
 \be\label{con-ex}
 H(f(t)|f_\infty(\theta, \kappa,\delta)) \le H(f(t=0)|f_\infty(\theta, \kappa,\delta)) \exp\left\{  - \frac{\delta^2}{\theta^\delta} \, t\right\}. 
  \ee
Exponential convergence in $L^1(\R_+)$ at the sub-optimal rate $\delta^2/2\theta^\delta$ then follows from the Csiszar--Kullback inequality \cite{Csi,Kul}, that reads
 \be\label{l1}
 \|f(t) - f_\infty(\theta, \kappa,\delta)\|_{L^1(\R_+)} \le 2 \sqrt{H(f(t=0)|f_\infty(\theta, \kappa,\delta)) }\exp\left\{  - \frac{\delta^2}{2\theta^\delta} \, t\right\}. 
 \ee

\section{Conclusions}
In this paper, we proved entropy-type inequalities satisfied by generalized Gamma densities, a class of skewed probability densities related to a variety of physical, biological and social phenomena. In particular, these inequalities allow to prove exponential in time convergence towards equilibrium in relative entropy  of the solutions to a class of Fokker--Planck type equations recently introduced in \cite{To3} to describe addiction phenomena in a multi-agent society of individuals.


\section{acknowledgement}
This work has been written within the activities of GNFM (Gruppo Nazionale per la Fisica Matematica) of INdAM (Istituto Nazionale di Alta Matematica), Italy.
The research was partially supported by the Italian Ministry of Education, University and Research (MIUR) through the ``Dipartimenti di Eccellenza'' Programme (2018-2022) -- Department of Mathematics ``F. Casorati'', University of Pavia  and through the MIUR  project PRIN 2017TEXA3H ``Gradient flows, Optimal Transport and Metric Measure Structures''.

\end{document}